# Haptic VR Simulation for Surgery Procedures in Medical Training

Zheng Jie Lim and Kian Meng Yap

*Research Centre For Human-Machine Collaboration (HUMAC), School of Engineering and Technology,*
*Sunway University, No. 5, Jalan Universiti, Bandar Sunway, 47500 Selangor Darul Ehsan, Malaysia.*
(Email: zhengjie.it@gmail.com, kmyap@sunway.edu.my)

**Abstract --- Traditional medical training faces challenges like ethical concerns, safety risks, and high costs. VR technology offers a promising solution but is limited by low complexity and lack of tactile feedback. This paper presents a cost-effective haptic VR surgery simulation which simulates realistic Kidney Transplant using commercial devices to enhance training authenticity and immersion. Trainees can conduct incision and anastomosis procedures using a haptic stylus device that provides tactile sensations. Results from the test with medical participants showed that haptic feedback positively enhances the VR medical training experience.**

## 1 INTRODUCTION

Traditional medical practices confront multifaceted challenges encompassing ethical concerns, safety hazards, limited accessibility, and escalating costs [1][2]. These issues impede the efficacy of training and the acquisition of surgical skills. Amidst these challenges, the integration of Virtual Reality (VR) technologies has emerged as a promising avenue to alleviate the aforementioned hurdles by providing a safe, controlled, and repeatable virtual environment for surgical training. However, despite its potential, several limitations such as high initial setup cost, low fidelity to replicate highly complex procedure task, and lack of tactile feedback hinder its effectiveness [3].

In this paper, we present a VR surgery simulation with haptic adoption. The objective is to provide a cost-effective and accessible simulation utilizing commercial haptic and VR devices, aiming to enhance the authenticity and immersion of current VR simulations. This approach is designed to cultivate surgical skills and improve the psychomotor performance of surgical trainees, with the long-term goal of producing competent and confident surgeons, thereby enhancing patient survivability and care.

## 2 SYSTEM ARCHITECTURE

The simulation is developed in Unity due to its versatility and flexibility in supporting a wide range of tools and applications including plugins necessary for haptic devices to communicate with the game engine [4]. To provide a specific example of a simulation designed to enhance surgical skills, a kidney transplant procedure is selected. This decision is motivated by the current lack of VR training for this procedure and its surging demand within the medical field [5, 6].

To effectively simulate the kidney transplant procedure, three primary types of actions are defined to encompass all procedural tasks: Trajectory Action for incision and suturing, Insert Action for placing the kidney, clamps, and retractors, and Remove Action for removing surgical tools. The development process is streamlined by utilizing the OramaVR MAGES SDK, which offers various features for creating action prototypes and setting up procedural flow using Scenegraph [7]. Consequently, the entire procedure, including the incision and closing of the abdomen, and the anastomosis of the kidney to the iliac vein, iliac artery, and bladder, can be accurately simulated.

The user (trainee) can view the simulation inside the computer using any commercial VR headset. They can control their movement and manipulate objects using the VR handheld controller in their left hand. Their right hand is equipped with the 3D Systems Touch haptic device, which includes a 6 degree of freedom stylus that mirrors the surgical tool in the virtual environment. The stylus is configured using Haptic Directs Unity Software

Development Kit (SDK) to produce tactile sensations when conducting incision and suturing in the simulation by replicating the dynamic tactile feel of surgical tools, such as scalpel when slicing through human organs. To determine the total force $F$ required to achieve similar bounciness and texture of real-life human organ, we proposed a haptic rendering method that uses an extended spring-damper model for the calculation as shown in Eq. (1).

$$F = -k(x - x0) - bv - \mu N - pt \quad (1)$$

Each human organ, represented by non-deformable virtual models, is configured using a distinct set of parameters in Eq. (1) to produce the desired level of haptic feedback. **k** represents the stiffness of the virtual organ, determining the elastic resistance based on displacement $(x - x0)$, where $x$ is the current position and $x0$ is the equilibrium position. $b$ is the damping coefficient for viscosity, representing the resistance to movement based on the velocity $v$ of the interaction point. $\mu N$ models the frictional resistance, where $\mu$ is the friction coefficient and $N$ is the normal force exerted by the stylus on the organ's surface. $pt$ is the pop-through force, which simulates the sudden change in resistance when penetrating tissue.

The complete setup of the hardware and software of the simulation is illustrated by Fig 1.

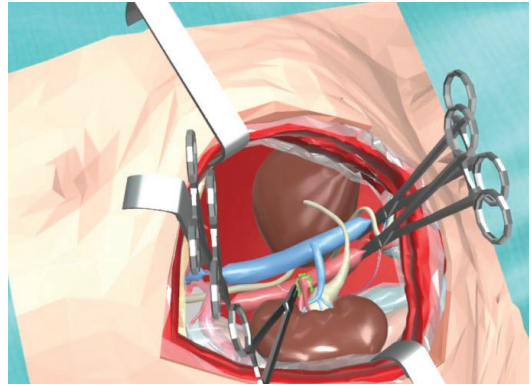

Fig.2 Anastomosis of Kidney Renal Artery to External Iliac Artery inside the simulation

## 3 TEST

The testing was conducted with 4 participants with medical background including surgery, orthopedic and human anatomy:

Stage 1: Participants were briefed on devices and the simulation. Then, the participants were given time to familiarize themselves with the devices until they are ready.

Stage 2: The participants would conduct the entire simulation using only VR devices from start to the end in a single session.

Stage 3: The participants would conduct the same simulation again using both VR devices and Touch Haptic Device.

Stage 4: Feedback from the participants is collected using a survey form.

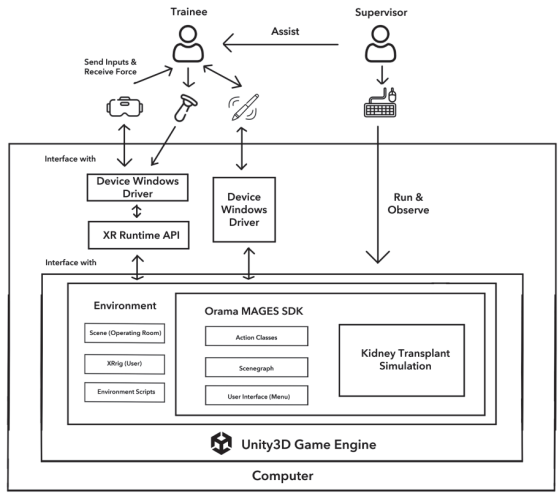

Fig.1 System Overview

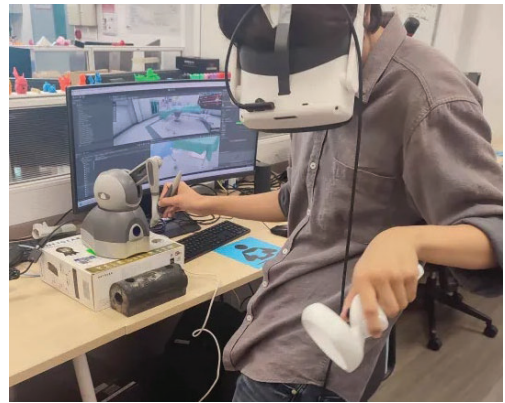

Fig.3 Participant Conducting Test

| Questions | 1 | 2 | 3 | 4 | 5 | Mean |
|---|---|---|---|---|---|---|
| C1. Integration of haptic feedback made learning in VR more fun and motivating. | 0 | 0 | 0 | 2 | 2 | 4.50 |
| C2. The haptic simulation has deepened my understanding of the body anatomy. | 0 | 0 | 2 | 2 | 0 | 3.50 |
| C3. The haptic simulation has built my confidence in conducting a real surgery procedure. | 0 | 0 | 3 | 1 | 0 | 3.25 |
| C4. The haptic simulation has improved my surgical skills or psychomotor skills (dexterity, hand-eye coordination). | 0 | 0 | 1 | 2 | 1 | 4.00 |
| C5. Similar experience can be achieved even if the VR simulation is conducted without any haptic feedback. | 2 | 2 | 0 | 0 | 0 | 1.50 |

Table 1. Ratings of the Participants on the Haptic VR Simulation at the Scale of 1 to 5 (Mostly Disagree to Mostly Agree)

## 4 Conclusion

Table 1 illustrates that the participants generally agree that haptic feedback positively enhances their VR training experience. Therefore, it can be concluded that the integration of haptic technology into VR training improves the overall effectiveness and realism of the simulation, leading to better skill acquisition and preparedness among surgical trainees. Additionally, we developed an effective haptic VR simulation leveraging accessible commercial devices, ensuring a cost-effective solution for widespread adoption in surgical training. However, the testing was limited by a small participant pool and lacks real surgical setting evaluations. Future work should expand the participant pool, include real surgical performance evaluations, and incorporate longer-term studies and diverse medical procedures to enhance the robustness of the findings.

## References

[1] R. Singh, R. Shane Tubbs, K. Gupta, M. Singh, D. G. Jones, and R. Kumar: "Is the decline of human anatomy hazardous to medical education/profession? — A review," *Surgical and Radiologic Anatomy*, vol. 37, no. 10, pp. 1257–1265 (2015)

[2] C. Brenna and S. Das, "Imperfect by design: the problematic ethics of surgical training," *Journal of Medical Ethics*, vol. 47, no. 5, pp 350-353 (2019)

[3] J. Pottle, "Virtual reality and the transformation of medical education," *Future Healthcare Journal.*, vol. 6, no. 3, pp. 181–185 (2019)

[4] C. Vohera, H. Chheda, D. Chouhan, A. Desai, and V. Jain, "Game Engine Architecture and Comparative Study of Different Game Engines," *2021 12th International Conference on Computing Communication and Networking Technologies (ICCCNT)*, Kharagpur, India, pp. 1–6 (2021)

[5] C. A. Mochtar et al., "Milestones of kidney transplantation in Indonesia," *Medical Journal of Indonesia*, vol. 26, no. 3, pp. 229–36 (2017)

[6] A. A. Kozan, L. H. Chan, and C. S. Biyani, "Current Status of Simulation Training in Urology: A Non-Systematic Review," *Research and Report in Urology.*, vol. 12, pp. 111-128 (2020)

[7] P. Zikas et al., "MAGES 4.0: Accelerating the World's Transition to VR Training and Democratizing the Authoring of the Medical Metaverse," *IEEE Computer Graphics and Applications*, vol. 43, no. 2, pp. 43–56 (2023)